\pdfoutput=1
\documentclass[aps,prd,amsmath,floats,floatfix, twocolumn,
superscriptaddress,nofootinbib,showpacs,longbibliography]{revtex4-1}
\usepackage[T1]{fontenc}
\usepackage[utf8]{inputenc}
\usepackage{lmodern}
\usepackage{verbatim}
\usepackage[dvipsnames, usenames]{xcolor}
\definecolor{linkcolor}{rgb}{0.0,0.3,0.5}
\usepackage[hypertexnames=false, unicode, colorlinks=true, linkcolor=linkcolor,
citecolor=linkcolor, filecolor=linkcolor,urlcolor=linkcolor,
pdfusetitle]{hyperref}
\usepackage[all]{hypcap}
\usepackage{graphicx}
\usepackage{xspace}
\usepackage{amssymb}
\usepackage{microtype}

\usepackage[english]{babel}
\usepackage{blindtext}

\usepackage[normalem]{ulem} 
\usepackage{bm} 
\usepackage{mathrsfs}

\graphicspath{
	{plots/}
}

\DeclareMathAlphabet{\mathpzc}{OT1}{pzc}{m}{it}

\usepackage[nomessages]{fp}

\begin{document}
\title{Interplay between numerical relativity and perturbation theory : finite size effects}
\newcommand{\UMassDMath}{\affiliation{Department of Mathematics,
		University of Massachusetts, Dartmouth, MA 02747, USA}}
\newcommand{\UMassDPhy}{\affiliation{Department of Physics,
		University of Massachusetts, Dartmouth, MA 02747, USA}}
\newcommand{\CSCVRUMass}{\affiliation{Center for Scientific Computing and Visualization Research, University of Massachusetts, Dartmouth, MA 02747, USA}}
\newcommand{\URI}{\affiliation{Department of Physics and Center for Computational Research, 
    University of Rhode Island, Kingston, RI 02881, USA}}    

\author{Tousif Islam}
\email{tislam@umassd.edu}
\UMassDPhy
\UMassDMath
\CSCVRUMass

\author{Gaurav Khanna}
\URI
\UMassDPhy
\CSCVRUMass

\hypersetup{pdfauthor={Islam et al.}}

\date{\today}

\begin{abstract}
We investigate the interplay between numerical relativity (NR) and point-particle black hole perturbation theory (ppBHPT) in the comparable mass ratio regime. In the ppBHPT framework, the secondary black hole is treated as a point particle, neglecting its finite size. Our study focuses on addressing the missing finite size effect in the ppBHPT framework and proposing a method for incorporating the size of the secondary into the perturbation theory framework. We demonstrate that by considering the secondary as a finite size object, the BHPT waveforms closely match NR waveforms. Additionally, we revisit the $\alpha$-$\beta$ scaling technique, which was previously introduced by Islam \textit{et al.}~\cite{Islam:2022laz}, as a means to effectively match ppBHPT waveforms to NR waveforms. We further analyze the scaling procedure and decompose it into different components, attributing them to various effects, including the corrections arising from the finite size of the secondary black hole.
\end{abstract}

\maketitle
\section{Introduction}
\label{Sec:Introduction}
Efficient detection and accurate characterization of gravitational wave (GW) signals from binary black hole (BBH) mergers require computationally cheap yet accurate multi-modal waveform models. The development of such models~\cite{Blackman:2015pia,Blackman:2017pcm,Blackman:2017dfb,Varma:2018mmi,Varma:2019csw,Islam:2021mha,bohe2017improved,cotesta2018enriching,cotesta2020frequency,pan2014inspiral,babak2017validating,husa2016frequency,khan2016frequency,london2018first,khan2019phenomenological} relies heavily on accurate numerical simulations of BBH mergers. The most accurate way to simulate a BBH merger is by solving the Einstein equations using numerical relativity (NR). Significant progress has been made in refining NR pipelines for BBH systems with comparable masses i.e. with $1 \le q \le 10$ (where $q:=m_1/m_2$ is the mass ratio of the binary with $m_1$ and $m_2$ being the mass of the primary and secondary black hole respectively) over the past two decades~\cite{Mroue:2013xna,Boyle:2019kee,Healy:2017psd,Healy:2019jyf,Healy:2020vre,Healy:2022wdn,Jani:2016wkt,Hamilton:2023qkv}. However, accurately simulating BBH mergers in the intermediate to large mass ratio regime ($10 \leq q \leq 100$) remains a challenging task. NR simulations in this mass ratio range become computationally demanding due to various factors. As a result, accurately modeling GW signals from BBH mergers in this regime using NR simulations becomes increasingly difficult.

On the other hand, point particle black hole perturbation theory (ppBHPT)~\cite{Sundararajan:2007jg,Sundararajan:2008zm,Sundararajan:2010sr,Zenginoglu:2011zz,Fujita:2004rb,Fujita:2005kng,Mano:1996vt,throwe2010high,OSullivan:2014ywd,Drasco:2005kz} provides an accurate modeling approach for extreme mass ratio binaries ($q \to \infty$). In ppBHPT, the smaller black hole is treated as a point particle orbiting the larger black hole in a curved space-time background. Substantial progress has been made in accurately simulating BBH mergers in this regime using the ppBHPT framework. However, as the mass ratio becomes less asymmetric and approaches the comparable to intermediate mass ratio regime, the assumptions of the ppBHPT framework start to break down. Consequently, the ppBHPT framework cannot generate accurate gravitational waveforms in this regime.

Understanding the interplay between NR and ppBHPT framework in the comparable to intermediate mass ratio regime is an active and exciting area of research in the field of gravitational waves. Several studies~\cite{Lousto:2010tb, Lousto:2010qx, Nakano:2011pb} have been conducted to investigate the limitations and accuracy of both approaches in this regime. 

Recently, significant advancements have been made in expanding the domain of NR and ppBHPT frameworks. These advancements include the development of the \texttt{BHPTNRSur1dq1e4} surrogate model \cite{Islam:2022laz} and a fully relativistic second-order self-force model \cite{Wardell:2021fyy}, as well as the expansion of numerical relativity (NR) techniques to simulate BBH mergers with higher mass ratios \cite{Lousto:2020tnb,Lousto:2022hoq,Yoo:2022erv, Giesler:2022inPrep}. The \texttt{BHPTNRSur1dq1e4} surrogate model, based on the ppBHPT framework, has demonstrated remarkable accuracy in predicting waveforms for BBH mergers in the comparable to large mass ratio regime. Through a simple but non-trivial calibration process called the $\alpha$-$\beta$ scaling, the ppBHPT waveforms are rescaled to achieve excellent agreement with NR data specifically in the comparable mass ratio regime. In parallel, the fully relativistic second-order self-force model has also shown promising results in accurately reproducing NR waveforms in the comparable mass ratio regime.

In this paper, we investigate the interplay between NR and ppBHPT framework in the comparable mass ratio regime through the lens of the $\alpha$-$\beta$ scaling. In particular, we focus on understanding the corrections due to the missing finite size of the secondary black hole in ppBHPT framework and proposing a robust method for incorporating the size of the secondary into the perturbation theory framework.
The remaining sections of the paper are structured as follows. Section~\ref{sec:BHPT} provides a concise overview of ppBHPT, the $\alpha$-$\beta$ scaling, and the perturbation theory framework that partially incorporates the size of the secondary black hole. 
In Section~\ref{sec:result}, we present our main findings and results.
To begin, Section~\ref{sec:radii} focuses on estimating the expected size of the secondary black hole. 
Next, in Section~\ref{sec:finite_size_on_wf}, we delve into the impact of considering the finite size of the secondary within the BHPT framework on the waveform. 
Section~\ref{sec:correct_blob_waveform} demonstrates the close agreement between the BHPT waveforms, incorporating the finite size secondary, and the NR waveforms. 
Finally, in Section~\ref{sec:alpha_beta_radii}, we revisit the $\alpha$-$\beta$ scaling and offer a detailed breakdown of its components. We attribute these components to different effects, including the corrections arising from the finite size of the secondary black hole.

\section{Black hole perturbation theory}
\label{sec:BHPT}
In this section, we present a concise summary of the perturbative techniques 
employed in this work to generate waveforms from BBH mergers in the comparable 
mass ratio regime. 
\subsection{Point-particle perturbation theory}
\label{sec:ppBHPT}
In the ppBHPT framework, the smaller black hole is modeled as a point-particle with no internal structure and a mass of $m_2$, moving in the spacetime of the larger Kerr black hole with mass $m_1$ and spin angular momentum per unit mass $a$.
Here, we provide an executive summary of this framework and refer to 
Refs. \cite{Sundararajan:2007jg,Sundararajan:2008zm,Sundararajan:2010sr,Zenginoglu:2011zz} for additional details.
First, we compute the trajectory taken by the point-particle and then we use that trajectory to compute the gravitational wave emission. 

During the initial adiabatic inspiral, the particle follows a sequence of geodesic orbits driven by radiative  energy and angular momentum losses.
The flux radiated to null infinity and through the event horizon are computed by solving the frequency-domain  Teukolsky equation\cite{Fujita:2004rb,Fujita:2005kng,Mano:1996vt,throwe2010high} using the open-source code \texttt{GremlinEq} \cite{OSullivan:2014ywd,Drasco:2005kz} from the Black Hole Perturbation Toolkit \cite{BHPToolkit}. The inspiral trajectory is then extended to include a plunge geodesic and a smooth transition region following a procedure similar to one proposed by Ori-Thorne \cite{Ori:2000zn}. We compute the transition between initial inspiral and the plunge using a generalized Ori-Thorne algorithm \cite{Hughes:2019zmt,Apte:2019txp}. Our trajectory model does not include the effects of the conservative or second-order self-force~\cite{Hinderer:2008dm}.

Once the trajectory of the perturbing compact body is fully specified, we solve the inhomogeneous Teukolsky equation in the time-domain while feeding the trajectory information 
into the particle source-term of the equation \cite{Sundararajan:2007jg,Sundararajan:2008zm,Sundararajan:2010sr,Zenginoglu:2011zz,Field:2021}.
This involves a four step procedure:
(i) rewriting the Teukolsky equation using compactified hyperboloidal 
coordinates is shown using standard Boyer-Lindquist coordinates that allow us to extract the gravitational waveform directly at null infinity while also solving the 
issue of unphysical reflections from the artificial boundary of the finite computational domain; (ii) obtaining 
a set of (2+1) dimensional PDEs by using the axisymmetry of the background Kerr space-time, and separating the 
dependence on azimuthal coordinate; (iii) recasting these equations into a first-order, hyperbolic PDE system; 
and finally (iv) implementing a high-order WENO (3,5) finite-difference scheme with Shu-Osher (3,3) time-stepping~\cite{Field:2021}.  
The point-particle source term on the right-hand-side of the Teukolsky equation requires some specialized techniques 
for a finite-difference numerical implementation~\cite{Sundararajan:2007jg,Sundararajan:2008zm}. We set the 
spin of the central black hole to a value slightly away from zero, $a/m_1 = 10^{-8}$ for technical 
reasons\footnote{For example, to avoid a change in the definition of the coordinates from Kerr to Schwarzschild.}. 

\subsection{Scaling between perturbation theory and numerical relativity waveforms}
\label{sec:ppBHPT_scaled}
While building a ppBHPT based waveform model for comparable to extreme mass ratio binaries Ref.~\cite{Islam:2022laz} introduced a simple but non-trivial scaling called the $\alpha$-$\beta$ scaling between ppBHPT waveforms and NR. It reads:
\begin{align} \label{eq:EMRI_rescale}
h^{\ell,m}_{\tt full, \alpha_{\ell}, \beta}(t ; q) \sim {\alpha_{\ell}} h^{\ell,m}_{\tt pp}\left( t \beta;q \right) \,,
\end{align}
where $\alpha_{\ell}$ and $\beta$ are determined by minimizing the $L_2$-norm between the NR and rescaled ppBHPT waveforms. 
Note that while $\alpha_{\ell}$ modifies the amplitude of the various spherical harmonic modes, $\beta$ changes the time/frequency evolution of the binary by rescaling the time.
After this $\alpha$-$\beta$ calibration step, the ppBHPT waveforms exhibit remarkable agreement with NR waveforms. 
For instance, in the comparable mass ratio regime, the dominant quadrupolar mode of the rescaled waveform agrees to NR with errors smaller than $\approx 10^{-3}$.

\subsection{Perturbation theory with finite size of the secondary black hole}
\label{sec:blob}
To incorporate the effect of a finite sized object in the context of the 
point-particle approach summarized above, we simply follow the method presented 
previously in Ref.~\cite{Barausse:2021}. In particular, we consider an extended 
object as a composition of point-particles arranged in a manner consistent with 
the shape and size of the object. As seen in Ref.~\cite{Barausse:2021} the dominant 
affect arises from the object's extent in the azimuthal $\varphi$ direction. In fact, 
the impact on the wave amplitude of the $h_{\ell m}$ mode can be computed by introducing 
an analytical factor as derived in Sec. IV in Ref.~\cite{Barausse:2021} for circular 
orbits of radius $r_0$. The factor $f(m)$ is given by 
\begin{equation}
  |f(m)|^2=2(1-\cos{\eta})/\eta^2\,,\quad \eta\equiv 2\pi m L/r_0\,.  
\end{equation}
Here $L$ represents the size of the extended object in the azimuthal direction and 
$m$ is the multipole mode. It is clear from the above expression that if the size 
of the object is small compared to the orbital radius, the factor has little effect. 
Therefore, the finite size correction is negligible during most of the slow inspiral 
phase of the binary evolution. However, this factor may deviate strongly from unity 
as the waveform peaks, i.e. when the orbital radius is close to the light-ring 
$r_0 = 3M$. 
\section{Results}
\label{sec:result}
In this section, we now provide a detailed discussion on how to associate a length scale to the secondary black hole in the BHPT framework and its effect on the waveform. Furthermore, we offer a comparison between the waveform of finite-sized BHPT and NR for various spherical harmonic modes.
\subsection{Estimating the size of the secondary black hole}
\label{sec:radii}
The horizon area $\mathcal{A}$ of a Kerr black hole (BH) with mass $m$ and spin $\chi$ is given by
\begin{equation}
\mathcal{A} = 8\pi \left( \frac{Gm}{c^2}\right)^2 (1 + \sqrt{1-\chi^2}).
\end{equation}
This enables us to associate a length scale with the black hole through its radius,
\begin{equation}
r_{\rm H} = \sqrt{\frac{\mathcal{A}}{4\pi}}.
\label{eq:radius}
\end{equation}

\begin{figure}
\includegraphics[width=\columnwidth]{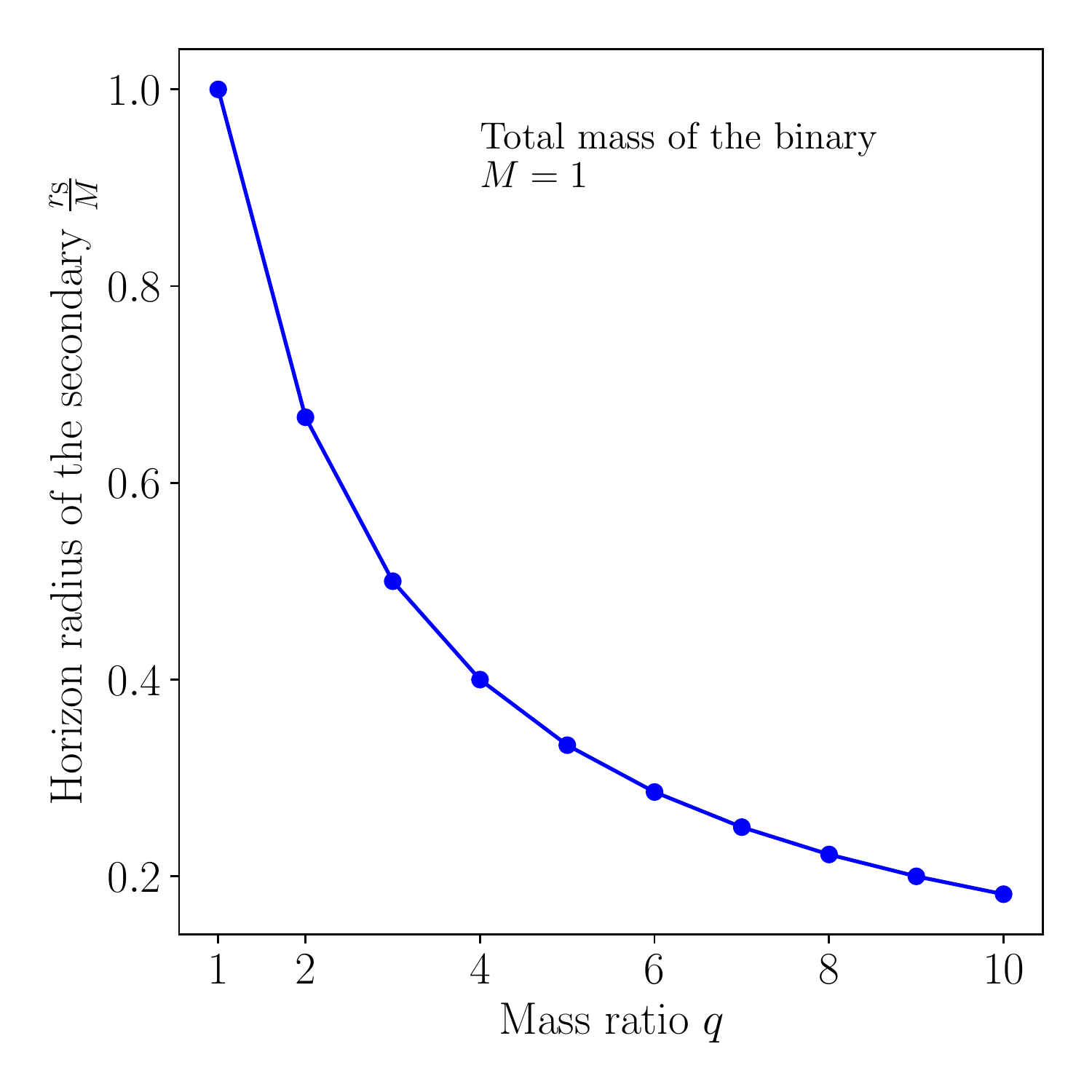}
\caption{We show the expected horizon radius of the secondary black hole, computed from Eq.(\ref{eq:radius}), as a function of the mass ratio $q$. More details are in Section \ref{sec:radii}.}
\label{fig:radii}
\end{figure}

In BBH merger simulations, such as in NR simulations, the total mass of the binary is typically normalized to unity in geometric units. The mass of the secondary BH is then determined as follows:
\begin{equation}
m_2 = \frac{1}{1+q}.
\end{equation}
Subsequently, we estimate the radius of the secondary BH for various mass ratios
\begin{equation}
r_s = \frac{2}{1+q}
\end{equation}
and present the results in Fig.~\ref{fig:radii}. It is worth noting that while the secondary BH in the ppBHPT framework accurately accounts for its mass, it is treated as a point particle and lacks a notion of radius. Conversely, NR simulations incorporate the size of the secondary BH.

\begin{figure}
\includegraphics[width=\columnwidth]{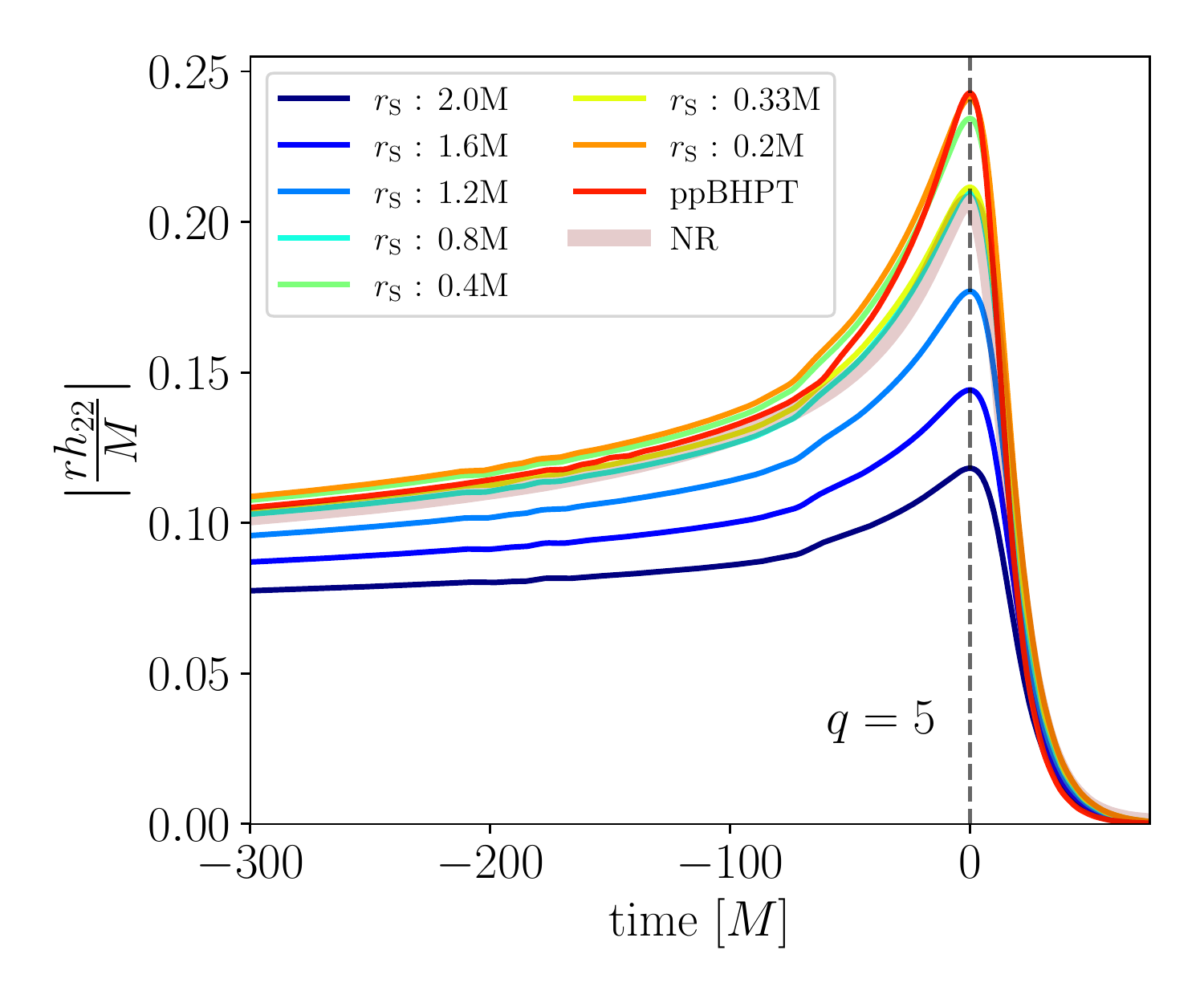}
\caption{We show the amplitudes of six BHPT waveforms at $q=5$ generated using a finite size secondary black hole. For comparison, we also include the amplitudes obtained from NR (thick red line) and ppBHPT (thin red line) waveforms. More details are in Section \ref{sec:finite_size_on_wf}.}
\label{fig:blob_q5}
\end{figure}

\subsection{Understanding the effect of finite size in ppBHPT waveforms}
\label{sec:finite_size_on_wf}
After proposing a notion for the expected radius of the secondary black hole, we now delve deeper into investigating the impact of the secondary's size on the waveform. Specifically, we choose $q=5$ and generate waveforms using the framework described in Section~\ref{sec:blob} for different sizes of the secondary BH. We simulate BBH mergers for six distinct values of the secondary's radius ($r_{\rm S}=[0.2,0.4,0.8,1.2,1.6,2.0]M$), while keeping the mass of the secondary constant. 

\begin{figure}
\includegraphics[width=\columnwidth]{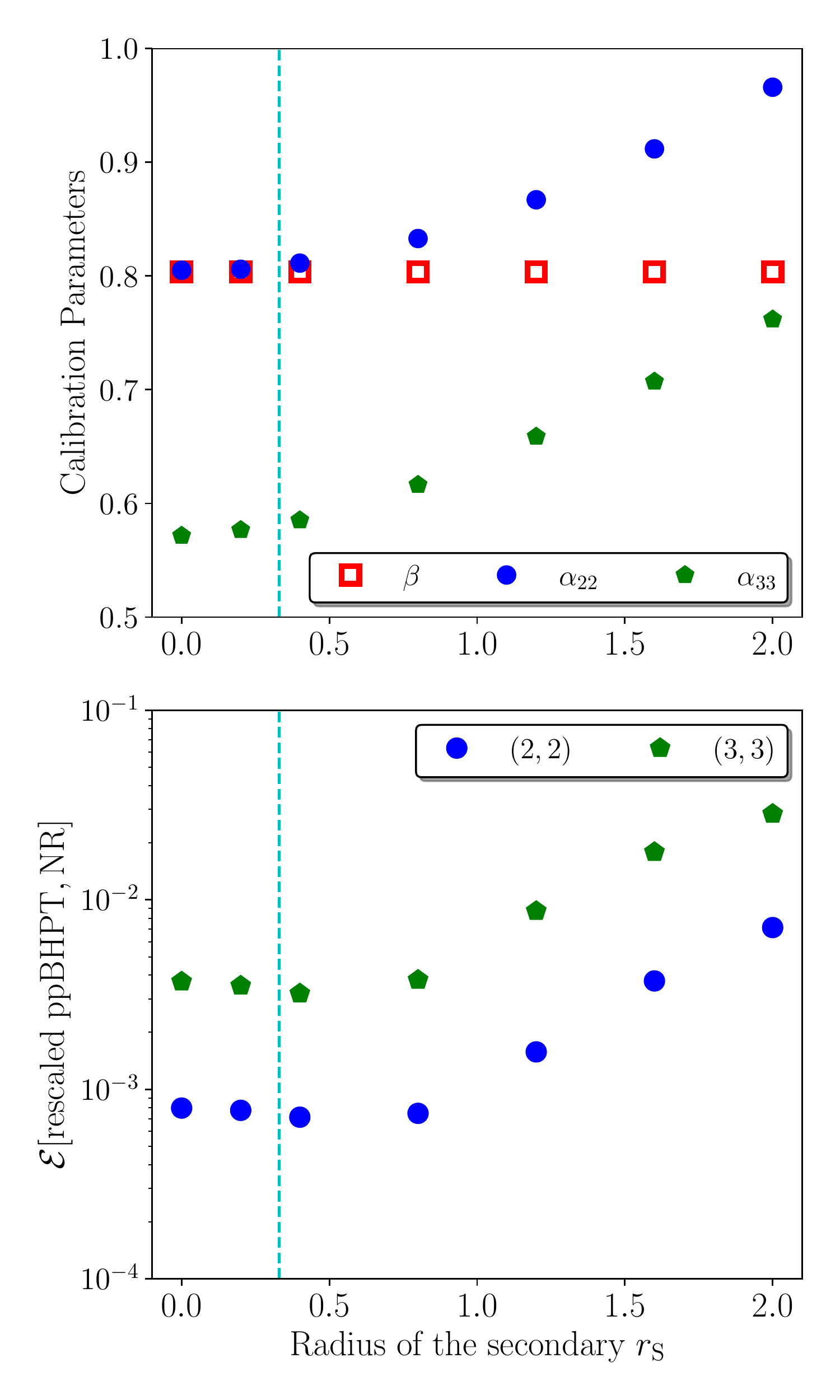}
\caption{We show the scaling parameters $\alpha_{22}$, $\alpha_{33}$ and $\beta$ as a function of the radius of the secondary black hole used in BHPT simulations (upper panel). Furthermore, we show the $L_2$-norm error between the rescaled BHPT waveform and NR after the $\alpha$-$\beta$ calibration for each simulation (lower panel). 
Dashed vertical lines denote the expected size of the secondary (i.e. $r_{\rm S}=0.33$) from Eq.(\ref{eq:radius}).
More details are in Section \ref{sec:finite_size_on_wf}.}
\label{fig:blob_alpha_beta}
\end{figure}

In Fig.~\ref{fig:blob_q5}, we present the amplitudes of the BHPT waveforms with varying sizes for the secondary BH. The waveforms are aligned such that the peak of the $(2,2)$ mode corresponds to $t=0M$. For comparison, we also include the amplitudes of the ppBHPT and NR waveforms. 
All the waveforms are scaled with the mass scale of $M$. To generate the NR waveforms, we utilize the \texttt{NRHybSur3dq8} waveform model~\cite{Varma:2018mmi}, which is a surrogate waveform trained on NR data hybridized with post-Newtonian-corrected effective-one-body inspirals. The use of \texttt{NRHybSur3dq8} introduces minimal systematic errors in our results, as the error in \texttt{NRHybSur3dq8} waveforms when compared to NR is close to the error between NR simulations with different resolutions~\cite{Varma:2018mmi}. Specifically, we extract the portion of the \texttt{NRHybSur3dq8} waveform corresponding to the last $\sim 5000M$ of the binary evolution, which corresponds to pure NR waveform data. We observe that the $(2,2)$ mode amplitude is initially the largest for the ppBHPT waveform. However, as we increase the size of the secondary BH, the amplitude gradually decreases. At a certain point, when the radius of the secondary becomes comparable to or larger than the expected radius for this mass ratio (which is $r_{\rm S}=0.33M$ as obtained from Eq.(\ref{eq:radius})), the amplitudes start to exceed those of the NR waveforms.

\begin{figure*}
\includegraphics[width=\textwidth]{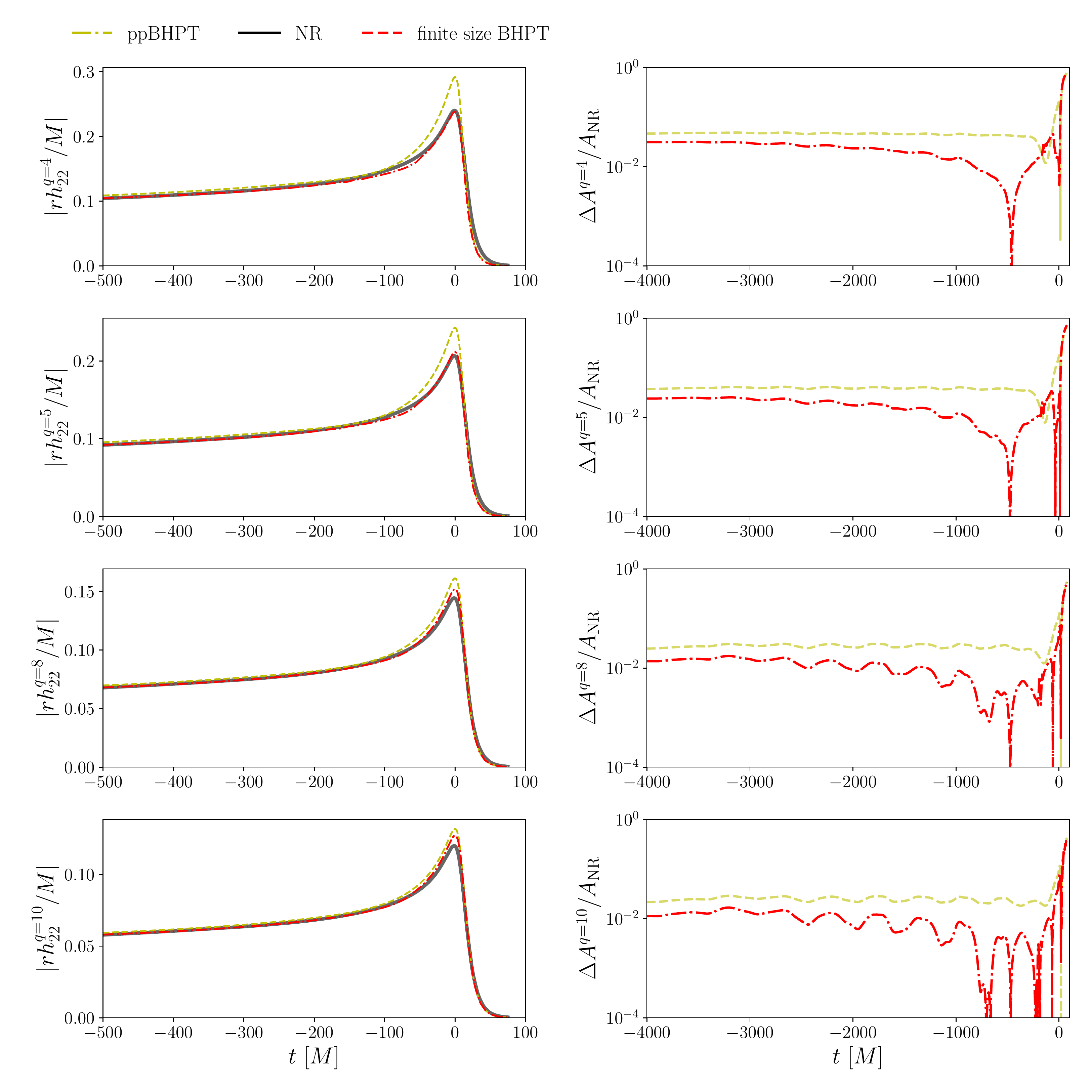}
\caption{We compare the amplitudes of the $(\ell,m)=(2,2)$ mode between finite size BHPT waveforms (red dashed lines), NR (solid black lines), and ppBHPT (yellow dashed lines) for mass ratios $q=[4,5,8,10]$ (left panels). Additionally, we present the corresponding relative errors with respect to NR (right panels). More details are in the text (Sec.~\ref{sec:correct_blob_waveform}).}
\label{fig:correct_blob_waveform}
\end{figure*}

Next, we proceed with the $\alpha$-$\beta$ calibration to the NR waveforms for each of the finite size BHPT waveforms generated. This allows us to determine the calibration parameters $\alpha_{\ell}$ and $\beta$ for the $(2,2)$ and $(3,3)$ modes. 
We should note that the finite-size BHPT framework, outlined in Section~\ref{sec:blob}, incorporates finite-size effects solely in the fluxes, while retaining the trajectory followed by the point particle. Consequently, these corrections affect only the amplitudes and do not impact the trajectory or the time/frequency evolution. This implies that although we may anticipate different values of $\alpha_{\ell}$ for the finite-size BHPT waveforms, the value of $\beta$ will remain unchanged as compared to the ppBHPT case.

In Fig.~\ref{fig:blob_alpha_beta}, we present the scaling parameters as well as the error between the rescaled BHPT and NR data after calibration, plotted as functions of the size of the secondary BH. As expected, the value of the scaling parameter $\beta$ remains constant regardless of the size of the secondary BH used in waveform generation.
Conversely, $\alpha_{\ell}$ varies as we change the size of the secondary BH. It is also worth noting that as the size of the secondary deviates from the expected value (i.e. $r_{\rm S}=0.33$), the error after calibration begins to increase. This presents an alternative approach to determine the correct size of the secondary by minimizing the error after calibration.

\begin{figure*}
\includegraphics[width=\textwidth]{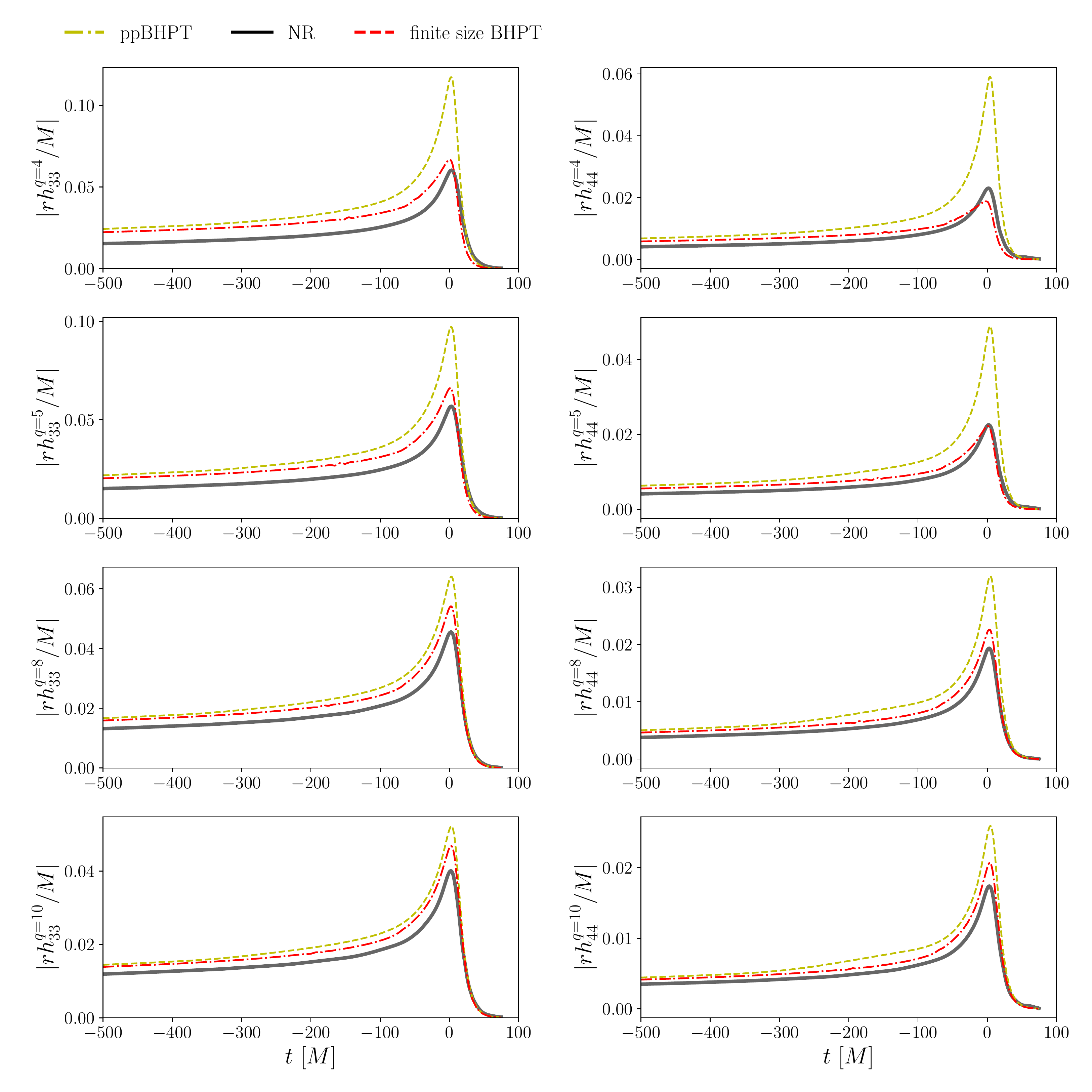}
\caption{We compare the amplitudes of the $(\ell,m)=(3,3)$ and $(\ell,m)=(4,4)$ mode between finite size BHPT waveforms (red dashed lines), NR (solid black lines), and ppBHPT (yellow dashed lines) for mass ratios $q=[4,5,8,10]$ (left panels). Additionally, we present the corresponding relative errors with respect to NR (right panels). More details are in the text (Sec.~\ref{sec:correct_blob_waveform}).}
\label{fig:correct_blob_waveform_33_44}
\end{figure*}

\subsection{ppBHPT waveforms with correct size of the secondary}
\label{sec:correct_blob_waveform}
It is promising to observe that assigning a finite size to the secondary black hole alters the waveform amplitude and could potentially offer an alternative method for matching BHPT waveforms to NR without performing any calibration. However, this approach relies on accurately determining the correct size of the secondary black hole from the outset. In this section, we explore whether it is feasible to establish such a framework. To begin with, we opt to employ the expected size of the secondary black hole, as estimated from Eq.(\ref{eq:radius}), as the input for our finite size BHPT simulation. It is somewhat encouraging to observe that, in Fig.~\ref{fig:blob_alpha_beta} for $q=5$, the error after the $\alpha$-$\beta$ calibration exhibits the minima very close to the expected radius obtained from Eq.(\ref{eq:radius}). 

We generate finite size BHPT waveforms for four different mass ratios: $q=[4,5,8,10]$. Subsequently, we compare these waveforms, particularly their amplitudes, with both NR\footnote{We use \texttt{NRHybSur3dq8} waveforms as a proxy of the NR data} and ppBHPT waveforms. In Figure~\ref{fig:correct_blob_waveform} (left panels), we present the $(2,2)$ mode amplitudes of the finite size waveforms, as well as those of the NR and ppBHPT waveforms. We observe that while ppBHPT waveforms consistently exhibit larger amplitudes than the NR data, the finite size BHPT waveforms closely match the amplitudes of NR waveforms for all the mass ratios examined in this study. To gain a deeper understanding of this agreement, we also plot the relative differences in the amplitudes compared to the NR data for both ppBHPT and finite size BHPT waveforms (right panels of Figure~\ref{fig:correct_blob_waveform}). Remarkably, we observe that the finite size BHPT waveforms consistently yield errors that are approximately one order of magnitude smaller than the errors between ppBHPT and NR waveforms. This clearly demonstrates the potential of the finite size BHPT framework in providing accurate waveforms in the regime of comparable mass ratios.

Next, we turn our attention to the higher order modes, specifically the $(3,3)$ and $(4,4)$ modes. We find that even with the inclusion of the expected size of the secondary in our BHPT simulations, the amplitudes of these higher order modes still exceed those of NR waveforms (Figure ~\ref{fig:correct_blob_waveform_33_44}). It is worth noting that the most significant impact of the finite size secondary is observed around the merger, where the amplitude of the finite size BHPT waveform decreases significantly. As expected, the effects of finite size corrections are more pronounced for smaller mass ratios, such as $q=4$, compared to larger mass ratios, such as $q=10$. 

\begin{figure}
\includegraphics[width=\columnwidth]{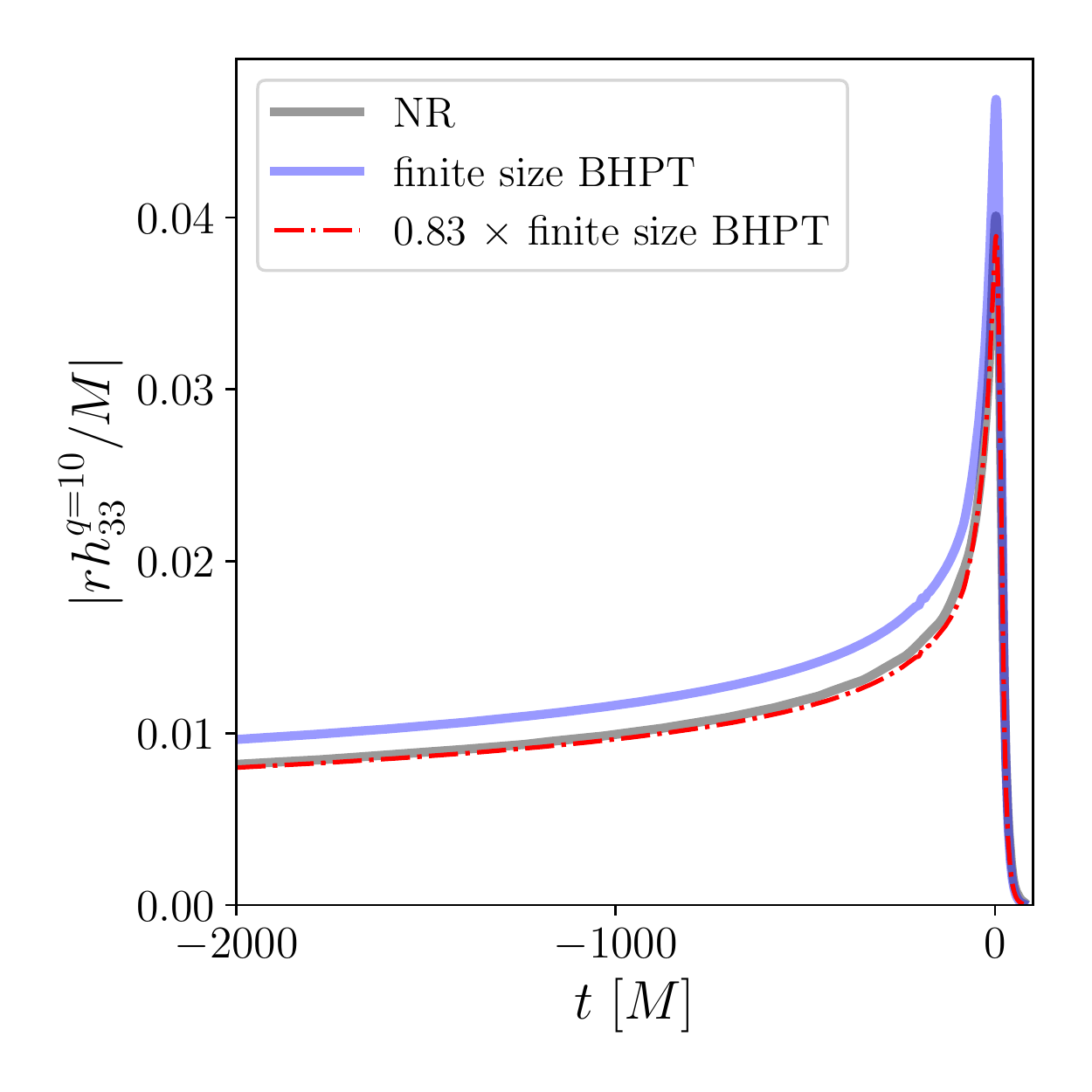}
\caption{We show the $(3,3)$ mode amplitudes of the NR and finite size BHPT waveforms for $q=10$ as well as the scaling between them. More details are in Section \ref{sec:correct_blob_waveform}.}
\label{fig:q10_33_rescaled}
\end{figure}

\begin{figure}
\includegraphics[width=\columnwidth]{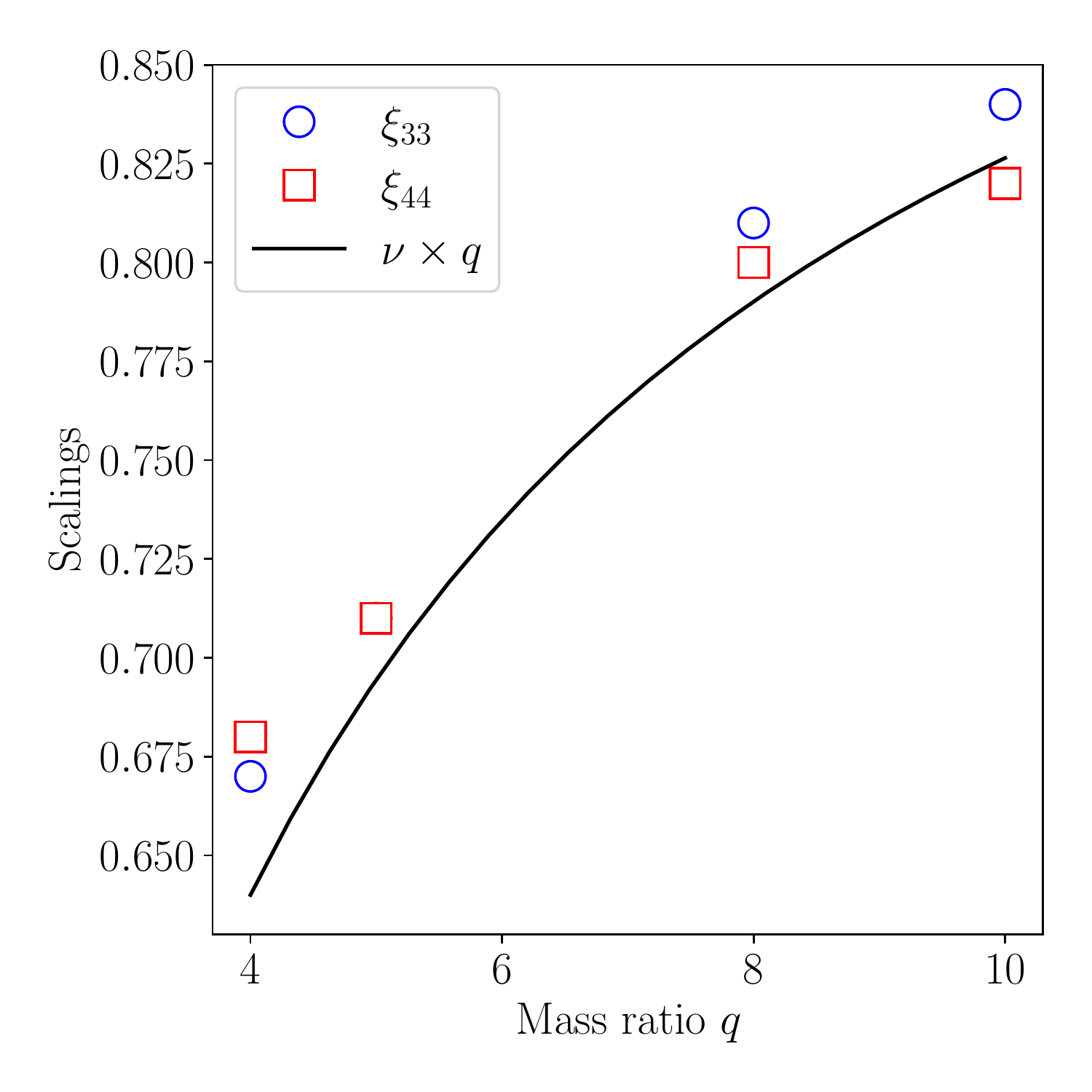}
\caption{We show the scaling factor $\xi_{\ell\neq 2,q}$, required to match a finite size BHPT mode amplitude to NR, as a function of the mass ratio $q$. For comparison, we also show the expected transformation factor ($\nu q$) between expanded in terms of $\frac{1}{q}$ and $\nu$. More details are in Section \ref{sec:correct_blob_waveform}.}
\label{fig:scalings_fsBHPT}
\end{figure}

We further observe that the amplitude ratio between the finite size BHPT waveform and NR waveform follows roughly similar scaling for both the $(3,3)$ and $(4,4)$ modes. Specifically, for a mass ratio of $q=10$, we find the following approximate relation:
\begin{align}
A_{33,\rm NR}^{q=10} \approx 0.81 \times A_{33, \rm fsBHPT}^{q=10},\\
A_{44,\rm NR}^{q=10} \approx 0.80 \times A_{44, \rm fsBHPT}^{q=10},
\end{align}
where $A_{33, \rm NR}$ and $A_{33, \rm fsBHPT}$ represent the $(3,3)$ mode amplitudes of the NR and finite size BHPT waveforms respectively. As an example, in Fig.~\ref{fig:q10_33_rescaled}, we show $A_{33, \rm NR}$ and $A_{33, \rm fsBHPT}$ as well as the scaling for $q=10$. We identify that the numbers $0.83$ and $0.81$ closely match the transformation factor $\nu q$ between waveforms expanded in terms of $\frac{1}{q}$ and $\nu$. For $q=10$, this transformation factor is 0.79. We generalize this relation as:
\begin{equation}
A_{\ell \neq 2,\rm NR}^{q} \approx \xi_{\ell\neq 2,q} \times A_{\ell\neq 2, \rm fsBHPT}^{q},
\end{equation}
where $\xi_{\ell\neq 2,q}$ is the scaling parameter required to match a finite size BHPT mode amplitude to NR. In Figure~\ref{fig:scalings_fsBHPT}, we present the scaling factors $\xi_{\ell,q}$ extracted for both the $(3,3)$ and $(4,4)$ modes across the four mass ratio values studied. Interestingly, we find that these scaling factors can largely be attributed to the simple scaling from the $\frac{1}{q}$ expansion to the $\nu$ expansion as mentioned earlier.

It is important to note that the finite size effect is smaller for the dominant $(2,2)$ mode. Consequently, the discrepancies in amplitudes observed in Figure~\ref{fig:correct_blob_waveform_33_44} or Figure~\ref{fig:q10_33_rescaled} may not be easily discernible for this mode. Therefore, when comparing waveforms generated with NR and BHPT frameworks, it becomes crucial to consider the higher order modes.

\begin{table*}[htb]
	\centering
	\setlength{\tabcolsep}{10pt}
	\begin{tabular}{l|l|l|l|l}
		\toprule
		$\ell$ &$A_{\alpha,\ell}$ &$B_{\alpha,\ell}$ &$C_{\alpha,\ell}$ &$D_{\alpha,\ell}$\\
		\hline
		2 &-1.330$\pm$0.007 &2.720$\pm$0.116 &-5.904$\pm$0.556 &5.548$\pm$0.833\\
        3 &-3.067$\pm$0.017 &6.244$\pm$0.265 &-9.944$\pm$1.261 &6.437$\pm$1.894\\
        4 &-3.909$\pm$0.032 &9.431$\pm$0.498 &-14.734$\pm$2.367 &9.744$\pm$3.556\\
        5 &-4.509$\pm$0.102 &4.751$\pm$1.554 &21.959$\pm$7.381 &-52.350$\pm$11.085\\
		\botrule	
	\end{tabular}
	\caption{Fitting coefficients  for $\alpha_{\ell}$ parameters as defined in Eq.(\ref{alpha_fit}).}
	\label{Tab:alpha_values}
\end{table*}

\begin{table*}[htb]
	\centering
	\setlength{\tabcolsep}{10pt}
	\begin{tabular}{l|l|l|l}
		\toprule
		$A_{\beta}$ &$B_{\beta}$ &$C_{\beta}$ &$D_{\beta}$\\
		\hline
		-1.238$\pm$0.003 &1.596$\pm$0.049 &-1.776$\pm$0.237 &1.0577$\pm$0.356\\
		\botrule	
	\end{tabular}
	\caption{Fitting coefficients for $\beta$ parameters as defined in Eq.(\ref{beta_fit}).}
	\label{Tab:beta_values}
\end{table*}

\section{Modelling the finite size effect}
\label{sec:alpha_beta_radii}
Based on the results presented in Section \ref{sec:correct_blob_waveform}, we now aim to revisit the $\alpha$-$\beta$ calibration procedure and break it down into smaller components to identify the effects of finite size.

In Ref.~\cite{Islam:2022laz}, fitting formulae for the calibration parameters $\alpha_{\ell}$ and $\beta$ have been developed for the $\alpha$-$\beta$ calibration procedure, as a function of the mass ratio $q$. These formulae read:
\begin{align}
\begin{split}
\alpha_{\ell}(q) =  1 & + \frac{A_{\alpha,\ell}}{q} + \frac{B_{\alpha,\ell}}{q^2}  + \frac{C_{\alpha,\ell}}{q^3} + \frac{D_{\alpha,\ell}}{q^4},\;
\end{split}
\label{alpha_fit}
\end{align}
and
\begin{align}
\begin{split}
\beta(q) =  1 & + \frac{A_{\beta}}{q} + \frac{B_{\beta}}{q^2} 
+ \frac{C_{\beta}}{q^3} + \frac{D_{\beta}}{q^4}.\;
\end{split}
\label{beta_fit}
\end{align}
Values of the fit coefficients are given in Table \ref{Tab:alpha_values} and Table \ref{Tab:beta_values} for $\alpha_{\ell}$ and $\beta$, respectively.

The calibration parameters $\alpha_{\ell}$ and $\beta$ account for two distinct effects in the context of the $\alpha$-$\beta$ calibration procedure: (i) the difference in mass scales between ppBHPT and NR; and (ii) the absence of finite size effects in the ppBHPT waveforms. Specifically, the mass scale in NR simulations is determined by the total mass of the binary $M$, whereas the mass scale used in ppBHPT corresponds to the mass of the primary black hole $m_1$. As a result, the transformation factor between the two mass scales can be expressed as $\frac{1}{1+1/q}$. Furthermore, the findings presented in Section~\ref{sec:correct_blob_waveform} suggest that even after incorporating the expected size of the secondary, the amplitude of the higher order modes needs to be rescaled by a factor that closely resembles a transformation between the $\frac{1}{q}$ expansion and the $\nu$ expansion. Furthermore, we observe that such a correction is unnecessary for the $(2,2)$ mode.

Based on the results presented in Section~\ref{sec:correct_blob_waveform}, we can decompose the $\alpha$-$\beta$ scaling into individual components and approximately determine the corrections resulting from finite size effects as:
\begin{equation}
\alpha_{\ell=2} = \frac{1}{1+1/q} \times \alpha_{\ell,\rm size}
\end{equation}
and
\begin{equation}
\alpha_{\ell \neq 2} = \frac{1}{1+1/q} \times  \left(\nu q\right) \times \alpha_{\ell,\rm size},
\end{equation}
where $\alpha_{\ell, \rm size}$ represent the finite size corrections in the amplitudes and need to be applied to the ppBHPT waveforms.

It is important to note that although we could not extensively investigate $\beta$ due to its unchanged nature under the finite size correction implemented in Section~\ref{sec:blob}, it exhibits similar qualitative and quantitative behavior as $\alpha_{\ell=2}$.
Hence, we express the potential finite-size correction in $\beta$ as:
\begin{equation}
\beta = \frac{1}{1+1/q} \times \beta_{\rm size},
\end{equation}
where $\beta_{\ell, \rm size}$ is the correction due to the missing finite size of the secondary in ppBHPT framework.

\begin{figure}
\includegraphics[width=\columnwidth]{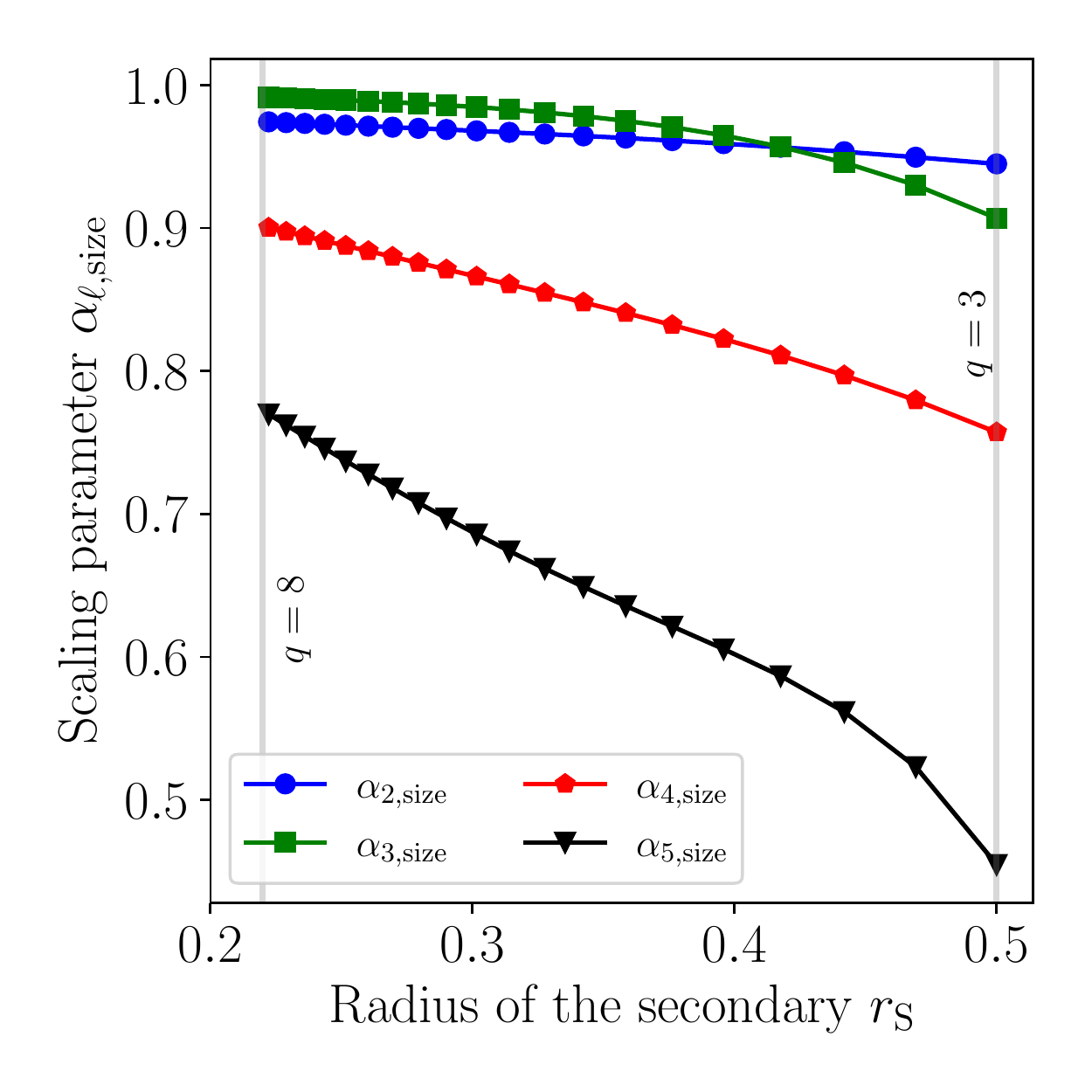}
\caption{We show the calibration parameters $\alpha_{\ell, \text{size}}$, which correct for the missing finite size effects, utilized in the $\alpha$-$\beta$ scaling of the \texttt{BHPTNRSur1dq1e4} model, as a function of $r_{\rm S}$, the expected radius of the secondary black hole. Vertical grey lines indicate $q=3$ and $q=8$. More details are in Section \ref{sec:alpha_beta_radii}.}
\label{fig:alpha_radii}
\end{figure}

We can establish a connection between the $\alpha_{\ell}$ and $\beta$ parameters (obtained from the fits presented in Eq.(\ref{alpha_fit}) and Eq.(\ref{beta_fit})) and the radius of the secondary black hole (obtained from Eq.(\ref{eq:radius})) for various mass ratios ranging from $q=3$ to $q=8$. 
This in turn gives us the connection between $\alpha_{\ell, \rm size}$ ($\beta_{\ell, \rm size}$) and the radius of the secondary (Fig.~\ref{fig:alpha_radii}).

This connection reveals the following behaviors:
\begin{align}
\alpha_{\ell=2,\text{size}} \approx & 0.9965083 -  0.1436961 \times r_{\rm S}\notag\\
&+0.2885172 \times r_{\rm S}^2 - 0.4157566 \times r_{\rm S}^3,
\end{align}
\begin{align}
\alpha_{\ell=3,\text{size}} \approx & 1.0661084 -  0.7671928 \times r_{\rm S}\notag\\
&+2.78573195 \times r_{\rm S}^2 - 3.77344743 \times r_{\rm S}^3,
\end{align}
\begin{align}
\alpha_{\ell=4,\text{size}} \approx & 1.04769829 -  0.98742265 \times r_{\rm S}\notag\\
&+1.98206353 \times r_{\rm S}^2 - 2.33710155 \times r_{\rm S}^3,
\end{align}
\begin{align}
\alpha_{\ell=5,\text{size}} \approx & 1.58061793 -  6.60322364 \times r_{\rm S}\notag\\
&+17.10135734 \times r_{\rm S}^2 - 16.75259602 \times r_{\rm S}^3,
\end{align}
and
\begin{align}
\beta_{\text{size}} \approx & 1.00082016 -  0.1298413 \times r_{\rm S}\notag\\
&+0.07899518 \times r_{\rm S}^2 - 0.05206764 \times r_{\rm S}^3.
\end{align}
It should be noted that smaller values of $\alpha_{\ell,\text{size}}$ correspond to larger corrections due to finite size effects. As anticipated, we observe that $\alpha_{\ell,\text{size}}$ decreases as the value of $q$ decreases, indicating that the finite size effect becomes more significant as the binary approaches an equal mass system. Additionally, the impact of the finite size effect is more pronounced in higher order modes compared to the dominant $(2,2)$ mode (Fig.~\ref{fig:alpha_radii}).

\begin{figure}
\includegraphics[width=\columnwidth]{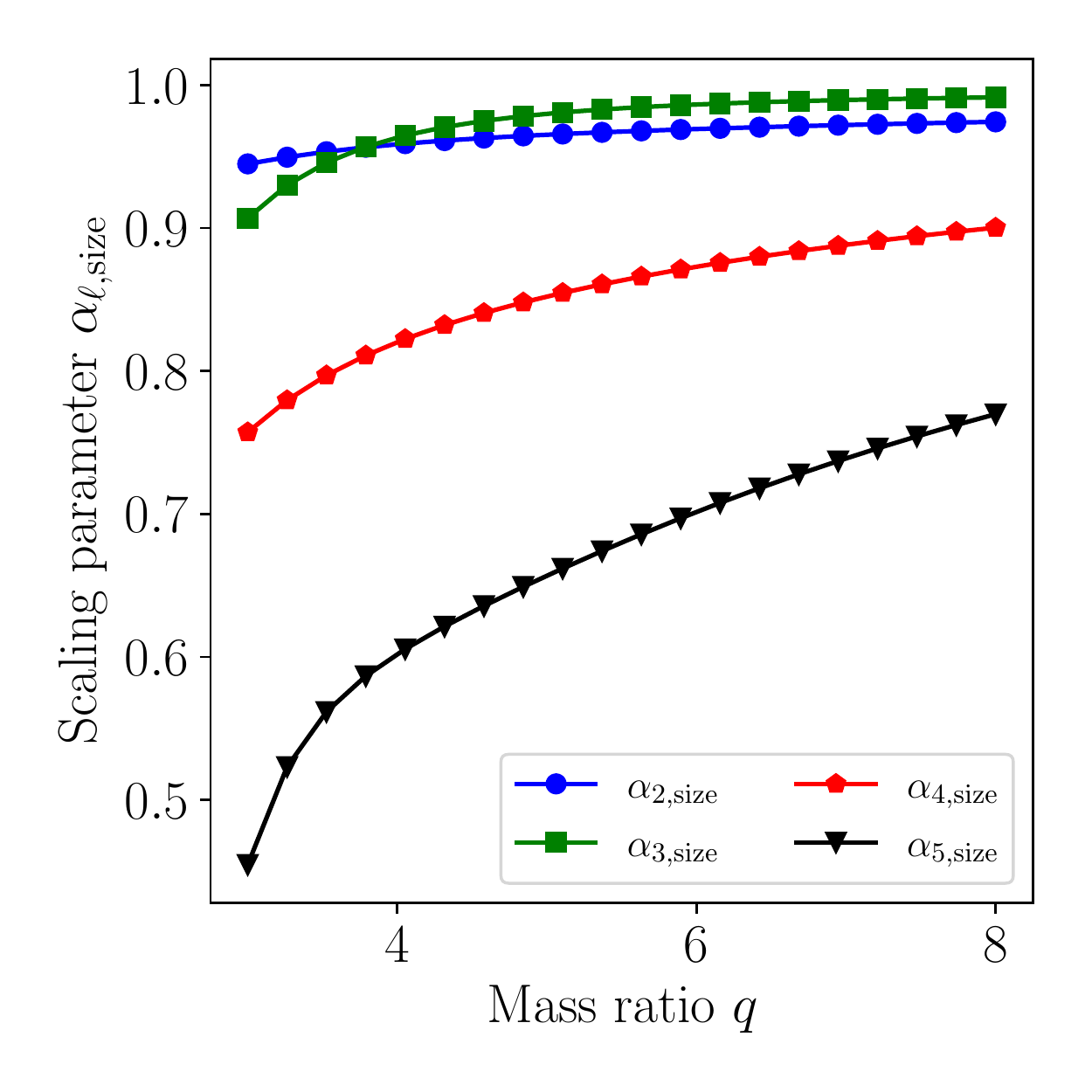}
\caption{We show the calibration parameters $\alpha_{\ell, \text{size}}$, which correct for the missing finite size effects, utilized in the $\alpha$-$\beta$ scaling of the \texttt{BHPTNRSur1dq1e4} model, as a function of the mass ratio $q$. More details are in Section \ref{sec:alpha_beta_radii}.}
\label{fig:alpha_q}
\end{figure}

Finally, we repeat the fitting in terms of $q$ and obtain:
\begin{align}
\alpha_{\ell=2,\text{size}} \approx & 0.98999937 -  0.14513987 \times (\frac{1}{q})\notag\\
&+ 0.22474979 \times (\frac{1}{q})^2 - 0.58826966 \times (\frac{1}{q})^3,
\end{align}
\begin{align}
\alpha_{\ell=3,\text{size}} \approx & 1.01037758 -  0.2900019 \times (\frac{1}{q})\notag\\
&+1.81636983 \times (\frac{1}{q})^2 - 5.63171863 \times (\frac{1}{q})^3,
\end{align}
\begin{align}
\alpha_{\ell=4,\text{size}} \approx & 1.01009073 -  1.1386684 \times (\frac{1}{q})\notag\\
&+2.65047754 \times (\frac{1}{q})^2 - 4.53281099 \times (\frac{1}{q})^3,
\end{align}
\begin{align}
\alpha_{\ell=5,\text{size}} \approx & 1.30116152 -  6.914990 \times (\frac{1}{q})\notag\\
&+26.51652 \times (\frac{1}{q})^2 - 40.08021 \times (\frac{1}{q})^3,
\end{align}
and
\begin{align}
\beta_{\text{size}} \approx & 0.9994668 -  0.2303172 \times (\frac{1}{q})\notag\\
&+0.3294111 \times (\frac{1}{q})^2 - 0.2743849 \times (\frac{1}{q})^3.
\end{align}

\begin{figure}
\includegraphics[width=\columnwidth]{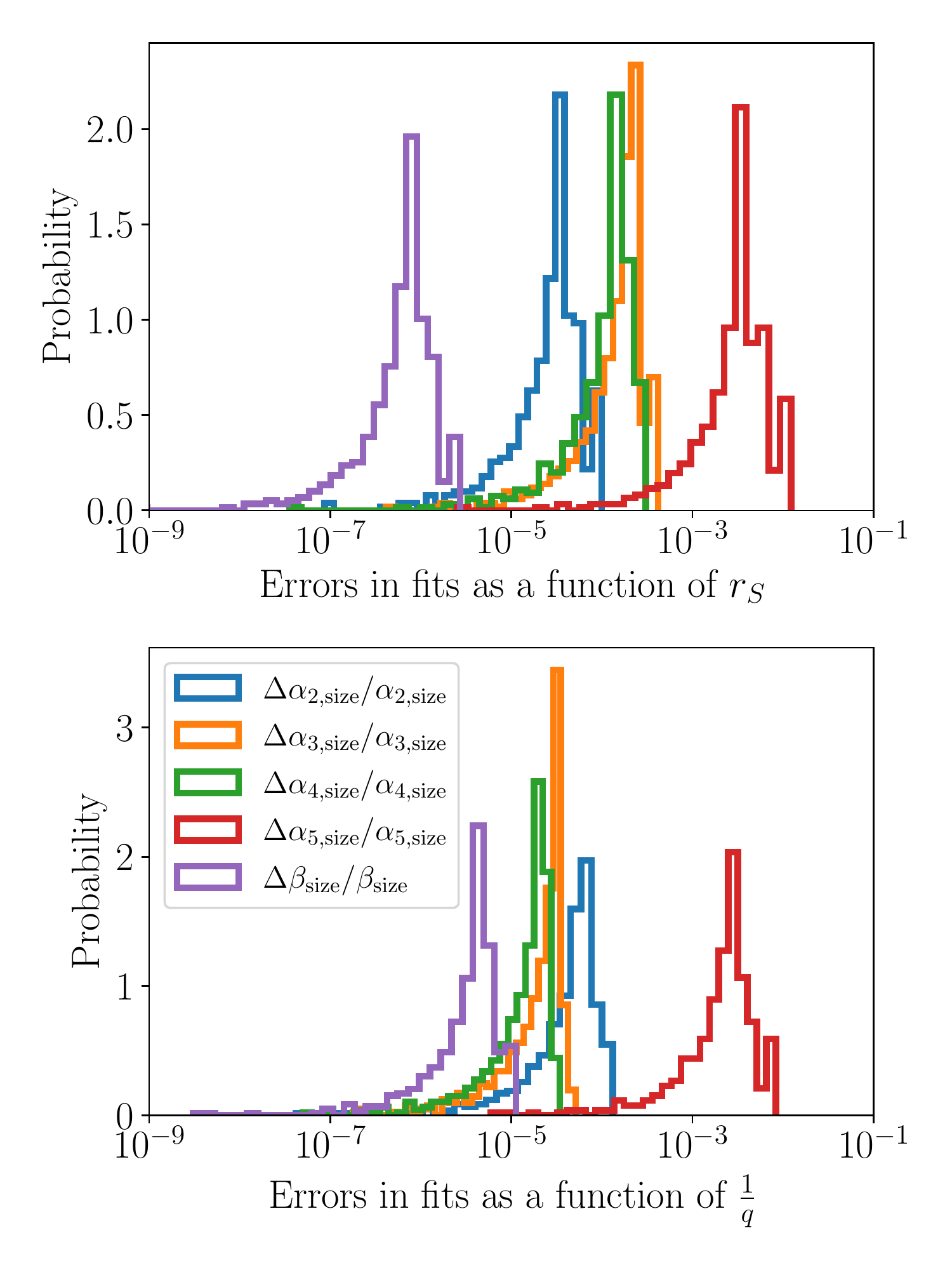}
\caption{We show the fitting errors in scaling parameters $\alpha_{\ell, \rm size}$ and $\beta_{\rm size}$ obtained by computing the difference between their numerical values and the values obtained from the analytical fits constructed as a function of both $r_S$ (upper panel) and $\frac{1}{q}$ (lower panel). More details are in Section \ref{sec:alpha_beta_radii}.}
\label{fig:alpha_beta_size_fit_err}
\end{figure}

We now compute the fitting errors in the scaling parameters $\alpha_{\ell, \rm size}$ and $\beta_{\rm size}$ (Fig.~\ref{fig:alpha_beta_size_fit_err}). To do this, we compare the numerical values of these parameters to the analytical fits constructed in this section. We evaluate the relative differences for 500 randomly chosen points in the mass ratio range of $q=3$ to $q=8$. The fitting errors are found to be very small, with values around $10^{-6}$ for $\beta$ regardless of whether the fits are based on $r_S$ or $\frac{1}{q}$. This demonstrates the reliability of the fits.

\section{Discussion \& Conclusions}
In this study, we have investigated the impact of finite size effects on BBH waveform modeling using the BHPT framework (described in Section~\ref{sec:blob}). We explored the behavior of the waveforms by incorporating a finite size for the secondary BH and compared them with both NR and ppBHPT waveforms. One key observation is that the introduction of a finite size for the secondary BH leads to a decrease in waveform amplitudes compared to the ppBHPT waveforms (Section~\ref{sec:correct_blob_waveform}; Figs.~\ref{fig:correct_blob_waveform},\ref{fig:correct_blob_waveform_33_44}). 

Additionally, we have performed the $\alpha$-$\beta$ calibration (proposed by Islam \textit{et al.}~\cite{Islam:2022laz}) to match finite size BHPT waveforms with NR waveforms (Section~\ref{sec:finite_size_on_wf}; Fig.~\ref{fig:blob_q5},~\ref{fig:blob_alpha_beta}). Interestingly, we found that $\alpha_{\ell,\text{size}}$ varies as the size of the secondary changes, indicating the specific impact of finite size effects on different multipole modes.

We further extracted the finite size corrections $\alpha_{\ell,\text{size}}$ and $\beta_{\text{size}}$ from the calibration parameters $\alpha_{\ell}$ and $\beta$ used to rescale a ppBHPT waveform so that it matches NR remarkably well in the comparable mass ratio regime. These corrections provide a systematic way to account for the missing finite size effects in the BHPT framework. We derived fitting formulas for $\alpha_{\ell,\text{size}}$ and $\beta_{\text{size}}$ as functions of the mass ratio as well as the expected radius of the secondary, which allows for a convenient determination of these corrections based on the physical properties of the binary system (Section~\ref{sec:alpha_beta_radii}; Figs.~\ref{fig:alpha_radii},~\ref{fig:alpha_q}).

Overall, we demonstrated that the amplitude of the finite size BHPT waveforms closely matches that of the NR waveforms for various mass ratios after efficiently incorporating finite size effects in the BHPT framework. This highlights the potential of accurately modeling BBH waveforms in the comparable mass ratio regime by considering the finite size of the secondary BH. However, further investigations are needed to explore methods for reliably determining the correct size in practical scenarios, as the correct size information is essential for achieving accurate waveform modeling.

It is worth noting that the finite-size BHPT framework (described in Section~\ref{sec:blob}; and in Ref.~\cite{Barausse:2021}) solely addresses the absence of finite size effects in the fluxes and amplitudes during the late inspiral-merger-ringdown phase of the waveform. Furthermore, it still relies on the trajectory of the point particle. While these limitations need to be addressed to develop a robust finite-size BHPT framework, it can already provide valuable insights into the dynamics of the binary and the intriguing interplay between NR and BHPT.

Our findings shed light on the interplay between NR and perturbation theory in the context of binary black hole waveforms. They underscore the significance of finite size effects and offer valuable insights into the potential of incorporating them into BHPT waveform models. By bridging the gap between perturbative approaches and NR simulations, these advancements enhance our understanding and modeling capabilities of GW signals from BBH mergers.

\begin{acknowledgments}
The authors acknowledge support of NSF Grants PHY-2106755, PHY-2307236 (G.K) and DMS-1912716, DMS-2309609 (T.I and G.K).  Simulations were performed on CARNiE at the Center for Scientific Computing and Visualization Research (CSCVR) of UMassD, which is supported by the ONR/DURIP Grant No.\ N00014181255 and the UMass-URI UNITY supercomputer supported by the Massachusetts Green High Performance Computing Center (MGHPCC). 
\end{acknowledgments}  

\bibliography{References}

\end{document}